\documentstyle[aps,prl,multicol,epsf]{revtex}

\begin{document}

\title{ Disorder-induced critical behavior in driven diffusive systems}

\author{Bosiljka Tadi\'c$^{*}$ }

\address{ Jo\v{z}ef Stefan Institute,  P.O.  Box 3000,
1001-Ljubljana, Slovenia} 

\maketitle
\begin{abstract}
\newline
Using dynamic renormalization group we study the transport  in
driven diffusive systems in the presence of quenched random drift
velocity  with long-range correlations along the transport direction.
In  dimensions $d\mathopen< 4$  we find  fixed points representing
novel universality classes of disorder-dominated self-organized
criticality, and a continuous phase transition at a critical variance
of disorder. Numerical values of  the scaling exponents characterizing
the distributions of relaxation clusters are in  good agreement
with the exponents measured in  natural river networks.
{\bf Published in: Phys. Rev. E, vol. 58, p. 168 (1998)}

\end{abstract}
\pacs{PACS numbers:  64.60.Lx, 05.60.+w, 64.60.Ak, 64.60.Ht}

\begin{multicols}{2}
\section{Introduction}
Various interacting driven systems self-organize into  critical steady
states, optimizing in that way their response as a {\it single
functional unit} \cite{Bak_book}. An important feature of these systems
is their response in the presence of disorder.
Effects of disorder   on critical properties  in steady states have
been  investigated \cite{Tetal},\cite{Toner} in cellular automata
and coarse-grained continuum models.
It has been recognized that disorder changes local relaxation rules and
breaks some symmetries of the dynamics, which may result in a
qualitatively different global dynamic state.
A distinct class of physical phenomena in driven systems exhibit
the scale-free behavior {\it only} in the presence of disorder.
Examples include energy transport in the integrate-and-fire oscillators
with diversity \cite{Al_diversity}, and Barkhausen noise in spatially
disordered ferromagnets \cite{BT_BN}. Fluctuations of optimal path in
heterogeneous materials \cite{optpath} and  landscape evolution due
to river networks flowing in naturally fractal environment
\cite{RN_experiment,RNs,DDRNs} also belong to this class of
dynamical systems.

Most studies of self-organized criticality (SOC) have been done on
sandpile automata, in which nonlinearity responsible
for SOC is due to threshold condition of toppling. In the  continuum
equation of motion for the dynamic variable height $h(\vec{x},t)$,
this leads to an infinite series of relevant operators
$\mu_n\partial^2h^n$ \cite{Diaz}.
Recent numerical simulations of stochastic automata with "soft" threshold
reveal  new universality classes of SOC and a  phase transition when
probability of toppling is varied \cite{LTU}.
Complementary to numerical simulations,
the renormalization group studies of continuum equations  are aimed
to characterizing the  critical behavior at large
distances and long times. Hwa and Kardar (HK) \cite{HK}  introduced
a transport equation which is compatible with all  symmetries
of anisotropic flow in open diffusive system, with leading nonlinearity
$\lambda \partial _\|h^2$ generated by nonlinear friction.

In this  work  we study the transport equation of  open
diffusive systems in the presence of quenched random drift velocity.
We adopt an anisotropic $d$-dimensional model for the height $h(\vec{x},t)$
transport with HK nonlinearity $\lambda \partial _\|h^2$,
and introduce  quenched disorder via a new term $p(x)\partial _\| h$,
which locally breaks the joint inversion  symmetry
$h\to -h$, $x_\| \to -x_\|$.  The symmetry is globally
restored by assuming the distribution of disorder with zero mean,
which thus excludes global current through the system.
We also consider long-range correlations of disorder
along the direction of transport varying with distance as
$\sim \gamma x_\|^{-2+\delta }$ and weak
(anti)correlation in perpendicular direction (see below).
In our model $p(x)$ represents spatially-varying local velocity
of profile fluctuations, which is motivated by  mass transport in
realistic granular and river flow  with a preferred direction
of drainage.
It should be stressed that our model differs from continuum models
of SOC studied so far  both in its symmetry properties and in
correlations of defects.
 Meanwhile, in  models of randomly driven
interfaces  \cite{RDIF} a similar disorder term  appears in  a
physically different context.

Using dynamic RG in the hydrodynamic (HD) limit we show that this type
of disorder represents a relevant perturbation in dimensions $d\le 4$,
leading to a new disorder-induced scaling  behavior. We calculate the
critical exponents at new  fixed points in the $\epsilon \equiv 4-d$
and $\delta$- expansion to leading order \cite{double-expansion}.
It is interesting to note that in the absence of disorder
the $\epsilon$-expansion to leading order yields the exact critical
exponents in the HD limit, as discussed in detail in Ref.\ \cite{HK}.
Using scaling arguments eligible for directed dynamic processes which
generate self-affine structures, we also determine the avalanche
exponents in terms of the anisotropy exponent.

The organization of the paper is as follows: In Sec.\ II we introduce
the stochastic differential equation with disorder and discuss
motivation for the long-range disorder correlations.
In Sec.\ III the details of
the dynamic renormalization group analysis are given. In Sec.\ IV we
discuss the critical behavior at disorder-induced fixed points for various
 physical values of the parameters and their  relevance for
the problems of river networks and strongly disordered dynamical systems.
 A short summary of the results is given in Sec.\ V.

\section{Stochastic equation of disorder-dominated networks}

We start with the anisotropic diffusion equation  for the height
transport \cite{HK} with nonlinear friction
\begin{equation}
{{\partial h}\over{\partial t}}= \nu _{\Vert } \partial
^2_{\Vert }h +\nu _{\bot }\partial ^2_{\bot }h - {{\lambda }\over{2}}\partial
_{\Vert }h^2 - p(x)\partial _{\Vert }h +\eta \ ,
\label{EOMO}
\end{equation}
and a time-dependent {\it nonconserving} Langevin force
\begin{equation}
\langle \eta (x,t)\eta (x^\prime t^\prime)\rangle = 2D\delta
^{(d)}(x-x^\prime)\delta (t-t^\prime )\ .
\label{noise}
\end{equation}
The new random  term proportional to $p(x)$ locally breaks joint
inversion symmetry, which is obeyed by the remaining terms \cite{HK}.  We
assume the distribution of $p(x)$ as
\begin{equation}
\langle p(x)\rangle _d =0 \ ; ~~~~~ \langle p(x)p(x^\prime
 )\rangle _d = \gamma f(x-x^\prime )
 \label{disorder}
\end{equation}
with $f(x) = x_\|^{-a}\ x_\perp ^c$.
For consistency of the perturbation expansion (see below) we
will chose $a=2-\delta $ and $c\sim O(2-z)$.
In Eq.\ (\ref{EOMO}) anisotropy signals the existence of a preferred
direction of mass flow, which is the subject of two nonlinear terms:
 $(\lambda /2)(\partial _\Vert h^2)$ and $\gamma f(x)(\partial _\Vert h)^2$.
The motivation for long-range disorder correlations is as follows:
 We assume that Eqs.\ (\ref{EOMO})-(\ref{disorder}) describe the evolution
of heights (e.g., of a  granular pile or landscape) which eventually leads
to a self-organized structured landscape with a network of
channels, along which the material is being eroded. It is important to
 keep in mind that these channels appear dynamically as a result
 of diffusion, which is influenced by an interplay of the above
 two nonlinear terms. Therefore, an initial configuration that is based
 mainly on the configuration of disorder, helps to imprint the channels by
setting locally most probable drift paths. However, since the system
is open and repeatedly perturbed by the nonconserving noise $\eta $,
the  once established network of  channels is likely to evolve under further
perturbations, reaching a new stationary configuration, in which effects of
disorder are altered.
An example of dynamically modified disorder effects can be
found in the field-driven random Ising model, in which pinning by local
 random fields appears weakened by sweeping an  avalanche of flipped spins
over certain pinning centers. The size of an  avalanche is the subject of
the dynamics itself. Another interesting example
is represented by erosion of natural landscapes due to water flow.
In the course of evolution, the originally preferred local  drift directions
 become uneffective at  sites  which are found inside the
correlated area  that already drains to a different direction.
Observations in  natural river basins reveal \cite{RN_observ,com_RN} a
 persistent correlation between the average soil slope at a site $x$ and
 area $s(x)$ that drains to that point as $\nabla h(x) \sim [s(x)]^{-1/2}$.
 Here the drainage
area $s(x)$ is  not fixed  but is determined self-consistently by the
dynamics itself. A nice example of this relation at work is  shown
in Ref.\ \cite{com_RN}, where a procedure  numerically iterated  to
{\it self-consistency} yields  a self-similar river network.
In the stationary critical state we have $s(x)\sim x_\|^{D_\|}\phi (x_\bot
/x_\|^\zeta )$, where $D_\|$ is the fractal dimension with respect to
parallel length, and $\zeta $ is the anisotropy exponent (see below).
It is reasonable to assume that for $x_\| \to \infty$ the scaling
function $\phi (r)$ behaves as a power of $r$, i.e., $\phi (r)\sim r^\eta$.
Therefore, for the {\it intermittent} dynamic regime [where  eroded
material from the area $s(x)$ is  accumulated at
point $x$ building up a shear stress $\sigma (x)$, and  erupting when the
stress exceeds a critical value $\sigma _c$], the leading nonlinear term is
proportional to $\sim (\nabla h)^2/x_\|^{D_\|-\eta \zeta }x_\bot ^\eta $.

In order to mimic the above processes in which the effects  of disorder are
 being dynamically modified, we only fix the disorder correlations in the
direction of transport. The transverse correlations are then
determined self-consistently by the  fluctuations in the critical steady
state \cite{c-ell}.
Notice that the difference $2-z$ is a  measure of the strength of critical
 fluctuations (see Sec.\ III).

\section{Dynamic renormalization group analysis}

The dynamic renormalization group consists of eliminating fast modes with
subsequent rescaling:
$t\to b^zt$,  $x_{\Vert } \to bx_{\Vert }$,  $x_{\bot }
\to b^{\zeta }x_{\bot }$, $h \to b^{\chi }h$,
where $z$, $\zeta $, and $\chi $ are the dynamic, anisotropy, and
roughness exponent,  respectively.
Naive dimensions of the coupling constants in Eq.\ (\ref{EOMO})
are then obtained from the scaled equation:
${{\partial h}/{\partial t}}= b^{z-2}\nu _{\Vert }
\partial ^2_{\Vert }h +b^{z-2\zeta }\nu _{\bot }\partial ^2_{\bot }h -
 b^{z+\chi -1}{{\lambda }\over{2}}\partial _{\Vert }h^2 + b^{z -1
+\mu _p}p(x)\partial _{\Vert }h +b^{z-\chi +\mu _{\eta }}\eta $,
where according to Eqs.\ (\ref{noise}-\ref{disorder})
we have  $\mu _p = -{{1}\over{2}}(a-c\zeta )$, and $\mu _{\eta }
=-{{1}\over{2}}\left[1+\zeta (d-1) +z\right] $.
In $(\vec{k},\omega )$-space the equation of motion becomes
\begin{eqnarray} \label{EOMk}
h(\vec{k},\omega ) = &&G_0(\vec{k},\omega )\left[\eta
(\vec{k},\omega )\right.\\ \nonumber
&&\left.- ik_\|{{\lambda }\over{2}} \int
{{d^dq}\over{(2\pi )^d}} {{d\omega ^\prime}\over{2\pi }} h(\vec{q},
\omega ^\prime )h(\vec{k}-\vec{q},\omega -\omega ^\prime )\right.\\ \nonumber
&&\left.-i\int {{d^dq}\over{(2\pi )^d}} {{d\omega ^\prime}\over{2\pi }}
(k_\|-q_\|)p(\vec{q})h(\vec{k}-\vec{q},\omega -\omega ^\prime )\right]
\end{eqnarray}
with the propagator $G_0(\vec{k},\omega )= \nu _\|k_\|^2 +\nu _\perp k_{\perp
}^2 -i\omega $.
Iterating Eq.\ (\ref{EOMk}) and eliminating fast modes
 leads to a diagramatic expansion (see Fig.\ 1).
 In the hydrodynamic (HD) limit
$k_\perp \to 0$, $\omega \to 0$, $k_\| \ll 1$, keeping  the lowest
respective orders of $k_\|$, we calculate the
one-loop contributions to the recursion relations [$\ell \equiv \log b$,
$S_d=2^{1-d}\pi ^{-d/2}/\Gamma (d/2)$ ]:
\begin{equation}
{{d\nu _{\Vert }}\over{d\ell }} = \nu _{\Vert }\left[z-2
+{{3\pi }\over{32}}S_du + 2\pi S_dw\right] \ ,
\label{nupar}
\end{equation}
\begin{equation}
{{dD}\over{d\ell }} = D\left[z-2\chi -(d-1)\zeta -1 +
{{\pi}\over{2}} S_dw\right] \ ,
\label{D}
\end{equation}
\begin{equation}
{{d\lambda }\over{d\ell }} = \lambda \left[{{z}\over{4}}(7-d)-
{{3}\over{2}} -{{3\pi }\over{4}}S_dw \right] \ ,
\label{lambda}
\end{equation}
\begin{equation}
{{d\gamma }\over{d\ell }} = \gamma \left[2z-2-a +c\zeta +\pi S_dw
\right] \; .
\label{gamma}
\end{equation}
Here the effective couplings $u$ and $w$ are found  to be
$u\equiv {{\lambda ^2D}\over{\nu _{\Vert }^3}}\left({{\nu
_{\Vert }}\over{\nu _{\bot }}}\right)^{{{d-1}\over{2}}}$  and  $w\equiv
{{\gamma }\over{\nu _{\Vert }^2}}\left({{\nu _{\Vert }}\over{\nu _{\bot
}}}\right)^{{{d-1}\over{2}}}$. A few comments are in order: $(a)$ As
usual in systems with quenched randomness, the perturbation expansion
is  made at fixed random noise $p(\vec{q})$, and subsequently the graphs
are averaged over the distribution of $p(\vec{q})$ leading to a dashed
line with a cross, which carries $\gamma q_\Vert ^{-a}q_\bot ^c$.
All graphs must be {\it connected} before this step is taken, thus leading
at most to one dashed-crossed line per loop. Due to the quenched nature of
the random noise [cf. Eq.\ (\ref{disorder})], a loop with dashed-crossed
line involves {\it no }frequency integration.  However,
averaging over the dynamic noise according to Eq.\ (\ref{noise}) leads to
a solid-circled line with two propagators and a factor $2D$, and an
integration over the internal frequency.
$(b)$ According to Eq.\ (\ref{EOMk}), the wiggly line
associated with the vertex $\lambda /2$ carries $ik_\|$, with $\vec{k}$
being the  momentum of the incoming line, whereas the wiggly line associated
with the vertex $p(\vec{q})$  carries  the momentum $i(k_\| -q_\|)$
of the outgoing line. Hence both graphs in Figs.\ 1a and 1b for the
renormalized propagator are proportional to $i(k_\|-q_\|)ik_\|$ and
thus do not contribue to the  vertex $\nu_\bot$. Therefore, we have
${{d\nu _{\bot }}/{d\ell }} = \nu _{\bot }\left[z-2\zeta \right]$,
leading to  $\zeta =z/2$. This argument is valid to all orders in the
HD limit.
$(c)$ Additional three graphs for $\lambda /2$ and $\gamma $ (not shown)
which are obtained by replacing the dashed-crossed line in Figs.\ 1d and 1f
with a solid-circled  line   (there are three such graphs
for for $\lambda /2$ and three for $\gamma$, corresponding to a circled line
along one of the three sides of triangle), however, give a null contribution
(same  as in Ref.\cite{HK}).
 Similarly, a contribution of the graph for
dynamic  noise $D$, which is obtained by replacing the dashed-crossed line in
Fig.\ 1c by a solid-circled line, vanishes.

On approaching a fixed point, we have from Eq.\ (\ref{D})
$\chi = \left[z(3-d)-2+ \pi S_dw\right]/4$. From Eqs.\ (\ref{lambda})
-(\ref{gamma}) we find
\begin{equation}
{{du}\over{d\ell }} = u\left[\epsilon -{{9\pi }\over{64}}S_du -
{{9\pi }\over{2}}S_dw \right] \ ,
\label{rec_u}
\end{equation}
\begin{equation}
{{dw}\over{d\ell }} = w\left[\delta - {{3\pi }\over{16}}S_du -
3\pi S_dw\right] \ ,
\label{rec_w}
\end{equation}
where the small expansion parameters are $\epsilon \equiv 4-d$ and
$\delta \equiv 2-a $,  and we have chosen $c=(2-z)/2 \equiv 1-\zeta$.
Notice that this choice of $c$ is selected by the structure of the
true expansion parameters $u$ and $w$, so that $\epsilon $  and $\delta $
appear as their anomalous dimensions, respectively. Also,
the disorder correlations $f(x_\| ,x_\bot )$ become isotropic
when $\zeta =1$, corresponding to the isotropic transport.
It should be stressed that for anisotropic disorder correlations no
additional parameters are generated to leading order.
Eqs.\ (\ref{rec_u})-(\ref{rec_w}) have four fixed points
$(u^\star,w^\star)$: (G) Gaussian\ $(0,0)$, (P) Pure
$\left({{64\epsilon }\over{9\pi S_d}}, 0\right)$, (R) Random
$\left(0, {{\delta }\over{3\pi S_d}}\right)$, and (M) Mixed
 $\left({{32(3\delta -2\epsilon )}\over{9\pi S_d}},
{{(4\epsilon -3\delta)}\over{9\pi S_d}}\right)$.

>From Eq.\ (\ref{nupar}) the dynamic exponent is obtained as
\begin{equation}
z= 2- {{3\pi }\over{32}}S_du^\star - 2\pi S_dw^\star \ .
\label{z_exp}
\end{equation}
Using Eq.\ (\ref{z_exp}) and the above scaling relations between $z$,
$\zeta $ and $\chi $, we find the exponents  in
the $\epsilon $- and $\delta $- expansion, which are shown in Table\ I.
Also shown are the exponents $\alpha $ and $\tau $ for the
probability distribution of  duration $P(t)\sim t^{-\alpha }$
and size of relaxation clusters, $P(s)\sim s^{-\tau }$,
which can be expressed in terms of  $\zeta $ using the following
scaling arguments (see also Ref.\cite{RNs}).
For {\it strictly directional} diffusion, the clusters can be
visualized as effectively planar structures with the  fractal
dimension  $D_\| =1+\zeta $.
The average size  of clusters scales as $\langle s\rangle \sim
L^{d_\ell}$, where  $L$ is the linear system size  and
$d_\ell =1$ for  the self-affine clusters (for which $\zeta \mathopen< 1$).
On the other hand,    $\langle s\rangle \sim L^{D_\|(2-\tau )}$,
and the scaling relation $D_\|(\tau -1) =\alpha -1 $ holds in the
steady state.  Using these relations  we find
$\tau = (1 + 2\zeta )/(1 + \zeta )$  and $\alpha = 1 + \zeta  $.

\section{Universality classes of disorder-induced criticality}

As seen from  Table \ I, the  fixed point (G) represents mean-field
SOC, which becomes   unstable for dimensions $d\mathopen< 4$
both with respect to  nonlinearity and disorder.
  Relative stability  of the other three fixed points in $(u,w)$-plane
 depends on the parameters $\epsilon$ and $\delta $ and on the initial
 value of the ratio $w/u$.  In Fig.\ 2 we show the
flow diagrams  of Eqs.\ (\ref{rec_u}-\ref{rec_w})
for $\delta =1$, $\epsilon=1$ and $\epsilon=2$.
(For convenience we use reduced
couplings $U\equiv \pi S_du/32$ and $W\equiv \pi S_dw$.)
In the  case $\epsilon=1$, competition between disorder  and the
$\lambda $-term leads to  two different types of  behavior, which  are
separated by the line $W/U =1$.
The fixed point (M), whose domain of attraction is the
line $W/U =1$,  is unstable in the direction perpendicular to the
critical line $W/U =1$, representing a phase transition from pure (HK)
to disorder-controlled SOC  with increasing variance of disorder.
We find qualitatively  the same  behavior  for
short-range  correlations ($\delta =2$) in $d=2$.
In the case of long-range correlated defects
($\delta =1$) in two dimensions (cf. Fig.\ 2)  the $(M)$
fixed point moves to the negative $U$-region
 and becomes spirally attractive.
The entire first quadrant flows towards the pure HK fixed point (P).
 The flow lines are first attracted to
a section of the curve connecting the (R) and (P) fixed points
approaching  fixed point (P)  under a nonzero angle.

Taking the  analytic continuation to  $\delta \to 1$ and $\epsilon \to$  1
or 2, corresponding to physical $d =$ 2+1 or 1+1 dimensions
\cite{DP-comment}, respectively, we obtain numerical values of the
exponents, which are listed in  Table\ II.
Here $\delta =1$ was taken as  a {\it typical} example of
long-range correlations. Notice that in contrast to $\epsilon $,
which is restricted to integer values, the parameter $\delta $ may
vary continuously in the range $0<\delta <2$. It is noteworthy that
the exponents at all three fixed points in $d=2+1$ are close to
values measured in natural river networks (RN). For river
basins around the world the exponents are found
\cite{RN_experiment} as: $\tau = 1.41-1.45$, $\alpha =1.67-1.92$,
$\zeta = 0.67 - 0.92$, and Hack's exponent $h=0.54-0.6$ satisfying
the scaling relation $h=1/\alpha $.  The roughness exponent
for large length scales \cite{chiex} was found in the range  $\chi =0.3-0.55$.
Variations in the values of the exponents depend on geographical
location where they have been  measured, and can be related to
locally dominated erosion mechanisms \cite{erosion}.
With regard to the results in Table\ II we would like to  point out
the following: ($i$) In the absence of disorder $\gamma =0$, corresponding to
the limit studied by Hwa and Kardar in Ref.\ \cite{HK}, the exponents are
within the range of the above RN exponents, indicating that the HK model of
flowing granular piles  captures the basic features of landscape evolution.
It should be stressed that in this case  ($\gamma =0$)
the values of the exponents are exact, and are  not a subject of higher-order
corrections in the perturbation expansion (see Ref.\  \cite{HK} for details).
($ii$) For finite disorder $\gamma \neq 0$ two more fixed points
 are accessible,  depending on the initial values of $\gamma $ and $\lambda $.
Therefore, variations of numerical values of the exponents
can be attributed to different universality classes, which are accessible
for varying initial strengths of disorder.
In particular, $\alpha $ decreases from 1.75 at HK limit
to 1.72 at (M) and  eventually to 1.66 at fixed point (R) by
increasing disorder $\gamma $ at fixed  $\lambda $ (see Fig.\ 2 top).
In addition, the exponents at fixed points  (M) and (R) vary with the
range of disorder  correlations $\delta $ (values of
$\delta $ in the interval $0<\delta <2$ correspond to long-range
correlations).
 For instance, for $\delta =1/3$, $\epsilon =1$ we
 have at (M) fixed point: $\zeta =0.83$, $\alpha =1.83$, and  $\tau =1.45 $.
It is interesting to note that the same values for cluster exponents
are obtained by the numerical procedure in Ref.\ \cite{com_RN}.
According to the discussion in Sec.\ II, we have that for $\zeta =0.83$
the exponent in the leading nonlinear term becomes
$D_\|-\zeta (1-\zeta )= 1.69 $, which is close to  $2-\delta =1.67$.
On the other hand, for shorter disorder correlations, e.g.,
for  $\delta =3/2$, we find $\alpha =1.64$ and $\tau =1.39$.
The importance of disorder for river networks has been also pointed out in
Ref.\ \cite{RPLEM}, where numerical simulations of  a cellular automaton
model of randomly pinned landscape evolution  yields the exponents very
close  to those at fixed point (M) (see Table \ II).
Moreover, in the absence of pinning, the same authors  \cite{RPLEM}
found the exponents close to HK fixed point (P) given in  Table\ II.

 At the (R) fixed point,
representing the universality class in strong disorder limit ($\lambda =0$),
the exponents are  in very good  agreement (cf. Table\ II) with the results
of numerical simulations of Ref.\ \cite{DDRNs}, suggesting small
higher-order corrections,
($d_\ell =0.98\pm 0.02$, $\zeta =0.66\pm 0.02$, and $\tau =1.40 \pm 0.02$)
for  self-affine networks with  disorder-dominated basins in two dimensions.
It has been argued in the literature \cite{optpath,DDRNs}
that the problem of optimal path in strongly disordered medium and Eden
growth processes also belong to this  universality class. In these systems
the disorder effects  are dynamically modified. Eden growth
is not defined as a disordered problem, however, an effective
 quenched  disorder with long-range correlations is self-generated
by the blocking effects of previously occupied sites (see second citation
in Ref.\ \cite{optpath}). Similarly, in the above mentioned example of
disorder-dominated basins in two dimensions sites that are already
connected at time $t$ influence the course of the  process at later time
steps. In  numerical simulations of  cellular automata models,  such as
done in Refs.\ \cite{DDRNs,RPLEM,LTU}, for instance,
a particular range of disorder correlations is {\it not specified}.
The exponents are measured in the emergent stationary critical state, which
is obtained after many successive updates. To our knowledge, a
stochastic differential equation for dynamically varying disorder effects
in these systems has not been considered so far.
Here we argue that Eq.\ (\ref{EOMO}) with disorder
correlations of the type $f(x)\sim x_\|^{-1}x_\bot ^{1-\zeta }$ might
capture the critical properties of these dynamical systems.

Following general scaling arguments of Ref.\ \cite{HK} we estimate
behavior of the order parameter, defined  in analogy to cellular
automata by the average  outflow current
\begin{equation}
< J(W)> =\int {{dt}\over{T}}\int
d^{d-1}x_\perp j(L_\|, x_\perp , t, W) .
\label{current0}
\end{equation}
In the steady state  $\langle J(W)\rangle \equiv 1$, and exhibiting
 fluctuations near the transition, where we have  $\langle J(W)\rangle
\sim W^\beta $.
For small disorder the local current is $j \sim h^2$, thus we have
  $j(L_\|,x_\perp , t, W) \sim b^{2\chi }j(b^{-1}L_\|,b^{-\zeta }x_\perp
, b^{-z}t, b^{-\mu _w}W)$ or
$j(L_\|,x_\perp , t, W) \sim x_\perp ^{2\chi /\zeta }\phi
(tx_\perp ^{-z/\zeta }, Wx_\perp ^{-\delta /\zeta })$.  Inserting into Eq.\
(\ref{current0}) and after extracting formally the $W$-dependence we find
$\beta = [2\chi +\zeta (d-1)]/\delta $.
The directed diffusion in our model represents some kind of a contact
process, therefore  for $t\to \infty $ and $x_\| \to \infty $
the following scaling relation holds:
  $\beta /{\bar{\nu }} =\alpha ^\prime $.  Here ${\bar{\nu}}  $
 is the parallel correlation length exponent,
and $\alpha ^\prime \equiv \alpha -1$ is the exponent of the
survival probability distribution. At (M) fixed point for $\delta =1$
and $\epsilon =1$ we find $\beta =0.78$ and ${\bar{\nu}} =1.08$.
Similar value  $\beta =0.86$ was found
in the stochastic cellular automaton (see first citation in Ref. \cite{LTU}).

\section{Conclusions}

We have demonstrated that our transport equation with
quenched disorder in the drift velocity with {\it anisotropic}
long-range correlations describes two novel universality classes of
critical behavior in open diffusive systems.
For finite disorder  we find SOC relevant for the scaling properties
of fractal river networks. For low disorder a crossover occurs to
the asymptotic
behavior controlled by the  HK fixed point \cite{HK}.
At critical variance of disorder  a continuous phase
transition occurs between the two different types of steady states:
channel-type flow for strong disorder and low friction, and surface-like
flow for low disorder and high friction.
Comparison of the numerical values of the avalanche exponents at
fixed points in 2+1 dimensions  with the exponents measured in  natural
 river networks \cite{RN_experiment}  is quite satisfactory.
Our analysis suggests that  natural river  networks may
result from the  interplay between quenched disorder and an effective
nonlinear friction. Variations in the range of disorder correlations
 $0<\delta <2$ appear as  a possible underlying mechanism which explains
the observed variations in the exponents of  natural networks.
A {\it distinct} universality class of disorder-induced
self-organized criticality is represented by the (R) fixed
point of our model, where $\lambda =0$. Evidence collected by
numerical simulations in Refs.\ \cite{optpath,DDRNs} suggest that
a number of other disordered dynamical systems should have the same
critical behavior described by the (R) fixed point.
In  the present work we pointed out the importance
of long-range correlations in this class of  self-organizing disordered
 systems.

\acknowledgements
This work was supported by the Ministry of Science and Technology of the
Republic of Slovenia.  I am grateful to Al Corral, Nicolay Antonov, Amos
Maritan and Achille Giacometti for helpful  discussions.

\narrowtext
\begin{table}
\caption{Scaling exponents at fixed points (R) and (M) to leading order in
 $\epsilon $- and $\delta $- expansion. Also listed are  mean-field exponents
 at (G), and exact HK results at (P) fixed point.}
\begin{center} \begin{tabular}{|c||c|c|c|c|c|}
~~~& $z $ & $ \zeta $ & $-\chi $ & $\alpha $ & $\tau $\\
\hline\hline
G & 2 & 1 & 1 & 2 & 3/2\\
\hline
P& ${{6}\over{7-d}}$& $ {{3}\over{7-d}}$& $ {{d-1}\over{7-d}}  $&
$ {{10-d}\over{7-d}}  $& $ {{13-d}\over{10-d}}   $\\
\hline
M &$2-{{3\delta +2\epsilon
}\over{9}}$ & $1-{{3\delta +2\epsilon }\over{18}}$ & $1-{{2\epsilon
}\over{3}}$ & $2-{{3\delta +2\epsilon}\over{18}}$& ${{3}\over{2}} -{{9\delta
+6\epsilon}\over{216}}$\\
\hline
R & $2-{{2\delta }\over{3}}$& $1-{{\delta }\over{3}}$
&$1-{{\delta }\over{4}} -{{\epsilon }\over{2}}$ & $2- {{\delta }\over{3}}$ &
${{3}\over{2}}-{{\delta }\over{12}}$\\
\end{tabular}
\end{center}

\narrowtext
\begin{table}
\caption{Numerical values of the critical exponents at various fixed points
for $\delta =$1, $\epsilon =1$ (upper part) and $\epsilon =2$.}
\begin{center}
\begin{tabular}{|c||c|c|c||c|c|c|}
~~~& $z $ & $ \zeta $ & $\chi $& $\alpha $ & $\tau $ &$\epsilon $\\
\hline\hline
P & 1.5& 0.75&
-0.5& 1.75& 1.428&~\\
M & 1.44& 0.72& -0.33& 1.72& 1.418&1\\
R & 1.33& 0.66& -0.25& 1.66& 1.40&~\\
\hline\hline P
& 1.2 & 0.6 & -0.2 &1.6& 1.375 &~\\
M &1.22& 0.62 & 0.33& 1.62 & 1.371 &2\\
R & 1.33& 0.66 &0.25 & 1.66&
1.40&~\\
\end{tabular}
\end{center}

\narrowtext

\begin{figure}
\epsfxsize=78mm\epsffile[10 206 558 620]{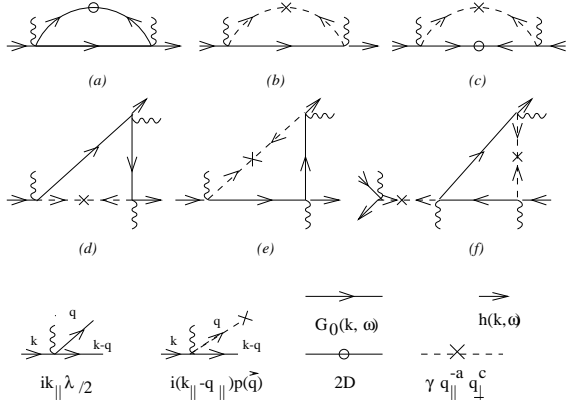}
\caption{One-loop diagrams with nonzero contributions to:
 (a,b) renormalized propagator,  (c) dynamic noise $D$, and vertices
(d,e) $\lambda /2$,  and  (f)  $\gamma $. The symbols are defined
in the bottom line.}
\label{fig1}
\end{figure}

\begin{figure}
\epsfxsize=78mm\epsffile[32 58 449 652]{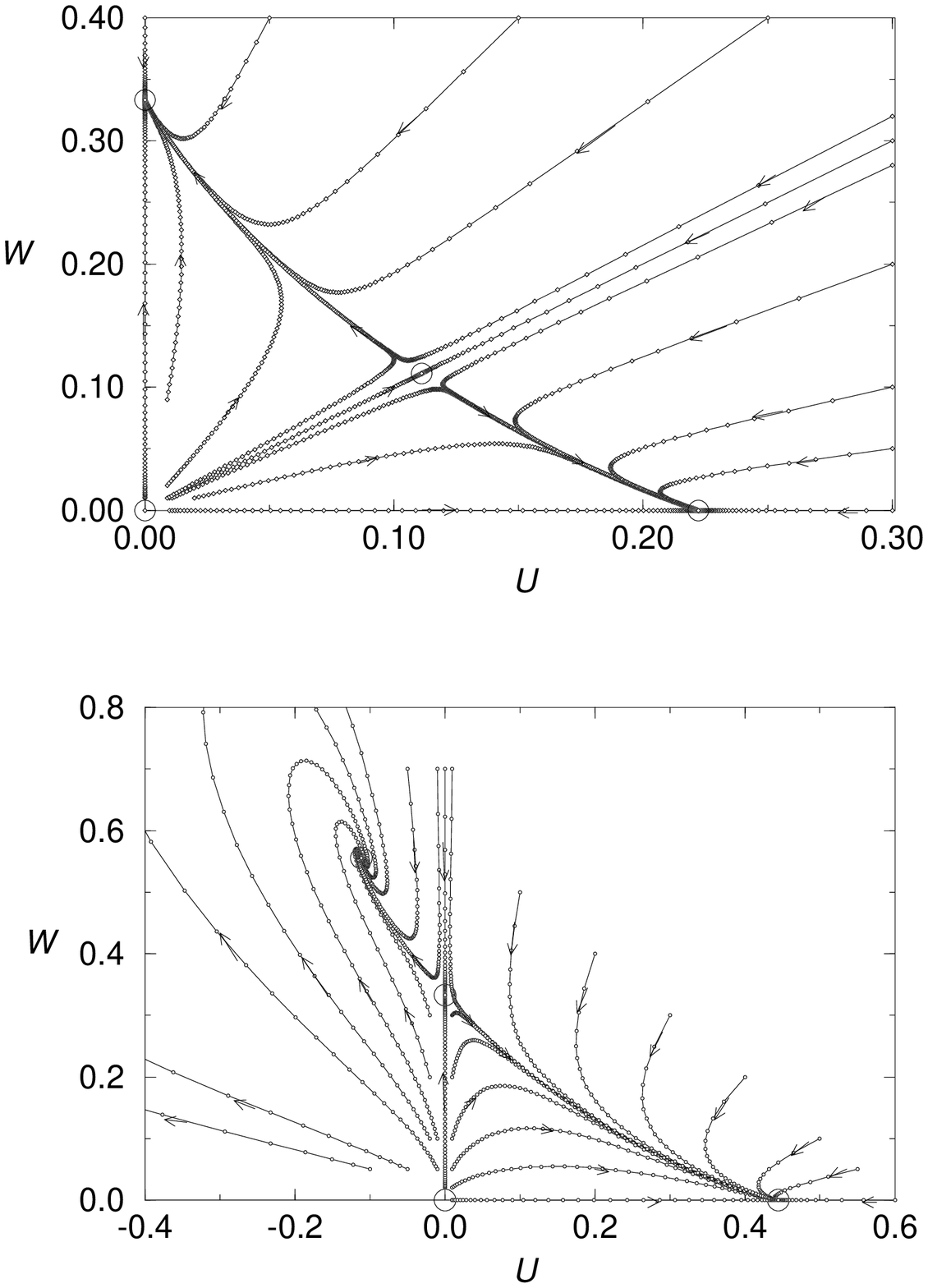}
\caption{Flow diagrams for $\delta =1$ and  (top) $\epsilon
=1$ and (bottom)  $\epsilon  =2$. Large circles represent
fixed points described in the text.}
\label{fig2}
\end{figure}

\end{table}
\end{table}
\end{multicols}


\begin{references}
\bibitem{c-ell} Therefore $c$ is renormalized together with other parameters
of the system $\lambda $ and $\gamma$, however, it remains of
$O(\epsilon ,\delta )$, i.e., in the same part of parameter space, thus
ensuring  a consistent perturbation expansion.
\bibitem{DP-comment}For dynamic
contact processes evolving on an internal time scale
$d$ can be identified as  $d-1$ spatial dimension and time, similar as
in directed percolation.
\end{references}
\end{document}